# Deterministic relation between optical polarization and lattice symmetry revealed in ion-doped single microcrystals

*Peng Li,[1,†,*] Yaxin Guo,[1,†] Ao Liu,[1] Xin Yue,[1] Taoli Yuan,[2] Jingping Zhu,[1] Yanpeng Zhang,[1] and Feng Li[1,*]*

[1]Key Laboratory for Physical Electronics and Devices of the Ministry of Education, School of Electronic Science and Engineering, Faculty of Electronic and Information Engineering, Xi'an Jiaotong University, Xi'an 710049, P. R. China

[2]School of Electronic Information and Artificial Intelligence, Shaanxi University of Science and Technology, Xi'an 710021, P. R. China

†These authors contributed equally
*Corresponding author: ponylee@stu.xjtu.edu.cn; felix831204@xjtu.edu.cn

**Abstract**: Rare-earth ions doped crystals are of great significance for micro-sensing and quantum information, whilst the ions in the crystals emit light with spontaneous partial polarization, which is, though believed to be originated from the crystal lattice structure, still lacking a deterministic explanation that can be tested with quantitative accuracy. We report the experimental evidence showing the profound physical relation between the polarization degree of light emitted by the doped ion and the lattice symmetry, by demonstrating, with unprecedented precision, that the lattice constant ratio c/a directly quantifies the macroscopic effective polar angle of the electric and magnetic dipoles, which essentially determines the linear polarization degree of the emission. Based on this discovery, we further propose a pure optical technology to identify the three-dimensional orientation of a rod-shaped single microcrystal using the polarization-resolved micro-spectroscopy. Our results, revealing the physical origin of light polarization in ion-doped crystals, open the way towards on-demand polarization control with crystallography, and provide a versatile platform for polarization-based microscale sensing in dynamical systems.

## 1. Introduction

Rare-earth ions doped crystals are widely applied in modern information technologies including displays,[1,2] microlasers,[3-5] solar cells,[6] optical sensing,[7-10] super-resolution imaging,[11-19] quantum processors,[20-29] etc. Unlike free atoms or ions, the optical properties of the doping ions are strongly influenced by the crystal field of the hosting material,[30-34] and a typical feature among these properties is the polarization of the emitted light by the ions, which is majorly determined by the crystal lattice symmetry. Such material effects are especially interesting in micro and nanoscale single crystals in which the light polarization, the crystal orientation and the lattice symmetry are closed linked. Indeed, it is well observed that the aligned assemble of microcrystals emits light of partially linear polarization while the emission from randomly oriented assemble remains unpolarized.[35-39] The partially polarized emission was further confirmed by the accurate micro-photoluminescence (μ-PL) measurements on single microcrystals in both real and angular spaces,[40-48] and a remarkable technology was recently developed to obtain the information of the three-dimensional (3D) orientation of a single microcrystal by just measuring the relative intensities of the polarization emissions towards the collecting lens.[49] With these advantageous properties, the micro- and nanocrystals can serve as powerful sensors for the single-particle tracking,[48] super-crystallographic reconstruction,[36] and indirect measurement of the local shear rate in microfluids.[10] Nevertheless, despite all the experimental evidence and practical applications, the physical origin of the polarization degree remains ambiguous. Although it is well accepted that the orientation of the electric and magnetic dipoles inside the crystal may be the source for the partial polarization of emission, the quantitative relation among the dipole orientations, the lattice symmetry and the emission polarization degree is still to be explored to fully understand the underlying physical mechanisms.

In this article, we report the experimental evidence that quantitatively reveals the profound relation between the polarization degree of the emitted light and the lattice structure, in a single crystalline microcrystal of hexagonal yttrium sodium fluoride doped with europium ions ($NaYF_4:Eu^{3+}$). We measure the linear polarization degree of the emitted light with unprecedented precision, from which we derive the orientations of electromagnetic dipoles based on the crystalline symmetry and the point group theory. Then we show that the obtained orientations of the dipoles are exactly oscillating along the composited vector of the c- and a-axes of the primitive cell, which defines the direction of a generalized crystal field. The link between the dipole orientations and the unit cell geometry quantitatively reveals the microscopic physical origin of the polarization degree of the emissions. We also obtain a complete set of mathematical formulas to identify the 3D orientations of single microcrystals from polarization resolved optical measurements. Our results provide an efficient method and a novel point of view for analyzing polarization related phenomenon in micro and nanoscale solids, and would facilitate the applications of ions-doped nanomaterials in the wide areas of sensing and information.

## 2. Experimental and analytical methods

The synthesized $NaYF_4:Eu^{3+}$ microcrystals crystallize in a standard hexagonal rod structure within the $P6_3/m$ space group, as confirmed by powder X-ray diffraction (see details in the Methods Section and **Figure S1**, Supporting Information). The microcrystals exhibit a uniform size of 8 μm in length and 2 μm in diameter, and are randomly dispersed on a quartz substrate to allow distinguishable single ones. We excite the in-plane single microcrystals using a focused laser beam with a wavelength of 395 nm, collect their photoluminescence (PL) from the top of (i.e., perpendicular to) the substrate by an objective lens (NA=0.8, 100×) and send to a spectrometer equipped with EMCCD for spectral analysis. We place half-wave, quarter-wave plates and a linear polarizer in front of the entrance slit for polarization analysis (see details in the Methods Section and **Figure S2**, Supporting Information). In order to eliminate the influence of the polarization preference of the spectrometer grating, the axis of the linear polarizer is always set parallel to the spectrometer entrance slit (defined as 0° and hereafter). We measure the linear polarization (resp. circular polarization) component of the photoluminescence by rotating the axis of the half-wave (resp. quarter-wave) plate in front of the linear polarizer. The results of the polarization-resolved spectroscopy are analyzed in two forms, using the Poincaré sphere (PS) and polarization fitting (PF).

In the PS form, we calculate three normalized Stokes parameters of ($S_1$, $S_2$, $S_3$) at six polarization bases

$$S_1 = \frac{I(0°)-I(90°)}{I(0°)+I(90°)} \quad (1)$$

$$S_2 = \frac{I(45°)-I(-45°)}{I(45°)+I(-45°)} \quad (2)$$

$$S_3 = \frac{I(\sigma^+)-I(\sigma^-)}{I(\sigma^+)+I(\sigma^-)} \quad (3)$$

where $I(\theta)$ [resp. $I(\sigma^\pm)$] is the emission intensity measured by the linear polarization component at angle $\theta$ (resp. the right or left circular polarization component). The linear polarization angle $\varphi$ and degree (LDOP) of the emission is derived by $2\varphi = arctan(S_2/S_1)$ and LDOP=$\sqrt{S_1^2 + S_2^2}$, respectively. In the PF form, $I(\theta)$ is measured from $\theta= 0°$ to $\theta= 360°$ with a step of 15°, which satisfies the equation for partially linearly polarized light:

$$I(\theta) = (A - B) \cdot cos^2(\theta - \phi) + B \qquad (4\text{-}1)$$

or

$$I(\theta) = A \cdot cos^2(\theta - \phi) + B \cdot sin^2(\theta - \phi) \qquad (4\text{-}2)$$

in which (A-B) and B are positive values representing the intensities of the purely linearly polarized and unpolarized components, respectively, and $\phi$ is the linear polarization angle. By fitting the experimental data of the $I$-$\theta$ relation with Equation (4-2), we can obtain the values of A, B and $\phi$. The linear polarization degree is then obtained by LDOP = $(I_{max} - I_{min})/(I_{max} + I_{min}) = (A - B)/(A + B)$.

We combine the two methods of display to reach the most precise analysis on polarization. The PS analysis allows to exclude the case of the non-zero circular polarization component (as will be proved experimentally later), which validates the usage of the PF method. Meanwhile, the PF allows higher accuracy of the obtained $\phi$ and LDOP. Therefore, combining the two methods allows very robust conclusions.

## 3. Results and Discussion

We first focus on the PL spectra of a single microcrystal within the wavelength range of 578-599 nm, with the spectra measured at 0°, 90°, ±45° and $\sigma^{\pm}$ polarization basis recorded in **Figure 1**a, where two typical emission bands of 579-585 nm and 585-595 nm are observed corresponding to the electric $^5D_1 \rightarrow {^7F_3}$ and the magnetic $^5D_0 \rightarrow {^7F_1}$ dipole transitions of the doped $Eu^{3+}$ ions, respectively.[50,51] The inset of Figure 1a shows the CCD image of the microcrystal whose axial direction (i.e., crystalline c-axis) exhibits a small angle with 0° (dashed yellow double arrow). As the electric $^5D_1 \rightarrow {^7F_3}$ transition is not well resolved spectrally, we analyze the magnetic $^5D_0 \rightarrow {^7F_1}$ transition, which is well resolved as three peaks labeled by I, II and III in Figure 1a. The physical origins of the three peaks are closed related to the crystal field.[32,34,52-54] As shown in Figure 1b, the energy levels of the $Eu^{3+}$ ions without the crystal background are created by the electron repulsion and spin-orbit coupling effects, leading to the energy levels $^5L_6$, $^5D_J$, and $^7F_J$. These levels are further split into fine structures by the crystal field related to the crystal symmetry, which can be viewed as an electrostatic perturbation on the electron could of the ions. The doped $Eu^{3+}$ ions are optically excited to the $^5L_6$ state from the ground state $^7F_0$, relax to the $^5D_0$ singlet by multiphonon channels,[55,56] and then to the $^7F_J$ multiplet by magnetic (J=1) and electric (J≠1) dipole optical transitions.[57-59] The triplet degeneracy of the J=1 state is completely lifted by the internal Stark effect induced by the crystal field, resulting in the split states $|-1\rangle$, $|+1\rangle$ and $|0\rangle$ corresponding to the emission peaks of I, II and III, respectively.[60]

The Stokes parameters $S_1$, $S_2$ and $S_3$ derived from the polarization-resolved PL of the magnetic $^5D_0 \rightarrow {^7F_1}$ transition are plotted in Figure 1c, where $S_3=0$ (i.e., zero circular degree of polarization) indicates that our single microcrystal host does not show chiroptical activity, different from several types of lanthanide-doped nanocrystals which are demonstrated to be with chirality.[61-64] The absence of chirality guarantees a clean link between light polarizations emitted by the ions and detected outside the crystal. Meanwhile, $S_1$ and $S_2$ display opposite values between the peaks I-II (585-590.5 nm) and the peak III (590.5-595 nm), implying orthogonal linear polarization. By fitting the spectrum as a sum of three peak areas of Lorentzian shape and calculating the integrated intensity of each, as shown in Figure 1d, we get the ($S_1$, $S_2$, $S_3$) values for the peaks I, II and III individually. The pure Lorentzian fitting matches perfectly the experimental data, thanks to the advantage of the excitation with narrow linewidth laser. The ($S_1$, $S_2$, $S_3$) of I, II and III are (0.49, 0.22, -0.01), (0.34, 0.15, -0.01) and (-0.68, -0.26, 0.03), all distributed in the equator plane of the Poincaré sphere, as shown in Figure 1e, confirming the absence of circular polarization degree for each individual peak. The linear polarization angle $\varphi$ and the LDOP are listed as sample No.1 in **Table 1**. We then use the PF method to obtain $\phi$ and LDOP. The spectrum is taken with a step of 15° as shown in **Figure 2**a, whist the fitting using Equation (4-2) is performed in Figure 2b. The obtained values of $\phi$ and the LDOP are listed as sample No.1 in **Table 2**. The very small differences in the results between the PS and the PF methods indicate that they are both valid for extracting the polarization information in the situation of $S_3=0$, and double confirm the accuracy of our measurements which is crucial for quantitative analysis.

The values of $\varphi$ or $\phi$ for the I and II (resp. III) are around 10° (resp. 100°), approximately the same (resp. complementary) angle of the crystal c-axis, indicating that the polarizations of I and II (resp. III) are parallel (resp. perpendicular) to the c-axis. The LDOP of all peaks are less than one, showing the nature of partially linearly polarization. We measured another four microcrystals of the same crystalline structure with different in-plane orientations (**Figure S3-S6**, Supporting Information), with the results summarized as samples No. 2-5 in Table 1 and 2. These samples show exactly the same features of polarization as sample No. 1, i.e., the orthogonal polarization between I, II and III and the same values of LDOP, confirming the robustness of our experimental conclusions.

The LDOPs of doped ion emissions are closed related to the properties of the corresponding transition dipoles inside the crystal. The $Eu^{3+}$ dopants, which replace the $Y^{3+}$ in the NaYF$_4$ crystals, occupy the crystallographic $C_s$ point groups.[60] The selection rules of this point group predict that there exist three peaks of the magnetic $^5D_0\rightarrow{}^7F_1$ transition corresponding to the I, II and III in our PL measurements, and also predict that two of the three originates from the oscillations of magnetic rotors and the other originates from the oscillations of magnetic dipoles (**Table S1,** Supporting Information).[34,65,66] It is obvious that the magnetic rotors lead to peaks I and II as they share the same polarization properties, while peak III are induced by the magnetic dipoles. To understand the collective behavior of all the magnetic dipoles and rotors in the microcrystal which essentially leads to the polarization features of the emitted light, we need first to be clear about the properties of a signal dipole or rotor, as illustrated in **Figure 3**a and b respectively. The linearly oscillating magnetic dipole radiates the linearly polarized $\sigma$ light, with its electric field $\vec{E}$ and magnetic field $\vec{H}$ oscillating perpendicular and parallel to the dipole orientation, respectively. The circularly oscillating magnetic rotor, which can be decomposed into two perpendicular magnetic dipoles with equal amplitudes and π/2 phase difference oscillating in the rotor plane, emits the linearly polarized $\pi$ light propagating within the rotor plane with its orthogonal $\vec{E}$ and $\vec{H}$ perpendicular and parallel to the rotor plane respectively, and, meanwhile, emits the circularly polarized $\alpha$ light propagating normal to the rotor plane with its $\vec{E}$ and $\vec{H}$ both parallel to it. The naming of $\sigma$, $\pi$ and $\alpha$ lights follows the previous definition in literatures.[34,52,53] The Y$^{3+}$ ions in the undoped NaYF$_4$ microcrystal constitute a compound hexagonal lattice as illustrated in Figure 3c. When replaced by the Eu$^{3+}$ emitters, the magnetic dipole and rotor is associated with each Eu$^{3+}$ site, showing a polar angle, noted as $\alpha$ for the dipole orientation and $\beta$ for the rotor plane, with respect to the crystalline c-axis. The hexagonal rotational symmetry, together with the uniformity of the crystal, requires that the absolute values of $\alpha$ or $\beta$ are the same for all Eu$^{3+}$ sites, as illustrated in Figure 3d. In reality, only a small portion of the Y$^{3+}$ is randomly replaced by the doped Eu$^{3+}$ emitter (5% doping level), which nevertheless should not affect the dipole or rotor orientations, as illustrated in Figure 3e. Herein, the dipoles and rotors are generally incoherent among different Eu$^{3+}$ emitters, unless reaching the condition of superradiance or stimulated emission which is very unlikely under our pumping scheme. The collective behavior of the dipoles and rotors leads to an effective macroscopic distribution of the hexagonal rotational symmetry, as shown in Figure 3f and g, since the dopant distribution can still be viewed as uniform in macroscale.

To link the polarization features of PL emission with the magnetic dipole and rotor orientations, we first analyze the emission of a single dipole or rotor detected by the experiment. We establish a Cartesian coordinate with z-axis being the crystalline c-axis, and x-axis being the direction along which the PL emission is collected by the objective. We start with the situation in which the dipole orientation is in the y-z plane, as illustrated in Figure 3h, where we only plot the electric field $\vec{E}$ which is actually detected by the EMCCD. To facilitate the analyzing process, we decompose the dipole moment to the y and z directions, forming two orthogonally orientated sub-dipoles. The energy flow of each sub-dipole electric field is then analyzed with its components along the x, y and z directions. We note $I_{ij}$ (i, j = x, y, z) as the light intensity polarized along i and propagating along j, as appears in Figure 3h, which covers the energy flow carried by the j-direction wavevector components of all electric fields. We define the total emission intensity of the magnetic dipole to be 2I, so that the sub-dipoles have the intensities of $2I \cdot cos^2\alpha$ and $2I \cdot sin^2\alpha$ according to Malus's Law. Due to the isotropy of the electric field in any plane perpendicular to the dipole orientation, we have $I_{xy} = I_{yx} = I \cdot cos^2\alpha$ and $I_{zx} = I_{xz} = I \cdot sin^2\alpha$. It is obvious that any dipole in Figure 3f can be viewed as the one in Figure 3h rotating around the z-axis by a certain angle $\vartheta$. Now we consider a total number of n randomly distributed $Eu^{3+}$ whose effective distribution of magnetic dipoles obeys rotational symmetry, then, as illustrated in Figure 3j in a top-down view (view along z direction), the integrated intensities associated with $I_{xy}$, $I_{yx}$ and $I_{zx}$ within the numerical aperture $\theta_{NA}$ of the collecting objective lens are

$$I_1 = \int_{-\theta_{NA}}^{\theta_{NA}} I_{xy} \cdot \frac{n}{2\pi} \cdot d\vartheta = \frac{nI\theta_{NA}}{\pi} \cdot cos^2\alpha \quad (5\text{-}1)$$

$$I_2 = \int_{-\theta_{NA}}^{\theta_{NA}} I_{yx} \cdot \frac{n}{2\pi} \cdot d\vartheta = \frac{nI\theta_{NA}}{\pi} \cdot cos^2\alpha \quad (5\text{-}2)$$

$$I_3 = \int_{-\theta_{NA}}^{\theta_{NA}} I_{zx} \cdot \frac{n}{2\pi} \cdot d\vartheta = \frac{nI\theta_{NA}}{\pi} \cdot sin^2\alpha \quad (5\text{-}3)$$

Seen from Figure 3f and j, $I_1$ and $I_2$ are purely linearly polarized perpendicular to the c-axis of the microcrystal (noted as $I_\perp$), while $I_3$ is purely linearly polarized parallel to it (noted as $I_\parallel$). According to Equation (5), we derive the relation between the polarized emission intensity and the polar angle of the magnetic dipoles:

$$\frac{I_\parallel}{I_\perp} = \frac{I_3}{I_1 + I_2} = \frac{tan^2\alpha}{2} \quad (6)$$

As already discussed before, the combined emission constitutes the magnetic dipole-induced peak III. When the polarization angle is either perpendicular or parallel to the c-axis, which is indeed the case in our experimental studies, the LDOP of the peak is expressed by

$$LDOP = \frac{I_{max} - I_{min}}{I_{max} + I_{min}} = \frac{I_\perp - I_\parallel}{I_\perp + I_\parallel} \quad (7)$$

according to the experimental results showing $I_\parallel < I_\perp$ for peak III (Figure 2b). By combing Equation (6) and (7), we derive a direct relation between LDOP and $\alpha$:

$$\frac{1-LDOP}{1+LDOP} = \frac{tan^2\alpha}{2} \quad (8)$$

using which the values of $\alpha$ calculated for the measurements of each sample are shown in Table 1 and 2. It is seen that the values display a very small fluctuation within ~3°, yielding an average of $\alpha$=30.2°. In addition, a significant rule we can derive from Equation (8) is that the numerical aperture of the objective lens, which initially appears in Equation (5) but cancels in Equation (6), doesn't influence the measured LDOP. We have confirmed this by performing experiments with another objective lens (NA 0.55, 100x) which yields the same LDOP. Therefore, the LDOP is solely determined by the polar angle $\alpha$.

We apply the same procedure for analyzing the incoherent emissions from the oscillating magnetic rotors, as illustrated in Figure 3i and k. Here, the five rotational components of $I_{1zy}$, $I_{zx}$, $I_{yx}$, $I_{xy}$, and $I_{2zy}$ can be collected in the objective lens, which can be divided into two groups: $I_{1zy}$, $I_{2zy}$ and $I_{zx}$ constitute $I_\parallel$; $I_{xy}$ and $I_{yx}$ constitute $I_\perp$. After the integration within the numerical aperture with $I_{1zy} = I_{zx} = I \cdot sin^2\beta$ and $I_{2zy} = I_{xy} = I_{yx} = I \cdot cos^2\beta$ (assuming the total emission intensity of the magnetic rotor to be 4I), the ratio between $I_\parallel$ to $I_\perp$ is

$$\frac{I_\parallel}{I_\perp} = \frac{1+2tan^2\beta}{2} \quad (9)$$

which further yields the relation between the LDOP and $\beta$ for the peaks I and II:

$$\frac{1+LDOP}{1-LDOP} = \frac{1+2tan^2\beta}{2} \quad (10)$$

with corresponding results also shown in Table 1 and 2. The average polar angles for $\beta_I$ and $\beta_{II}$ are 59.4° and 52.0°, respectively. Again, the LDOP is merely determined by $\beta$, irrelevant to the numerical aperture of the objective lens.

Besides the magnetic dipoles and rotors, the $C_s$ point group also allows the existence of electric dipoles and rotors, which show linear and rotating electric oscillations, respectively, as illustrated in Table S1 (Supporting Information). These includes the $^5D_0 \rightarrow ^7F_0$ and $^5D_0 \rightarrow ^7F_2$ transitions at the wavelength ranges of 576-578 nm and 608-632 nm. The detailed measurements and analysis related to the electric dipoles and rotors are shown in **Figures S7-S9** and **Table S2** (Supporting Information), from which we obtain the relation between the LDOP and the polar angles

$$\frac{1-LDOP}{1+LDOP} = \frac{2}{tan^2\alpha} \quad (11)$$

$$\frac{1+LDOP}{1-LDOP} = \frac{2}{1+2tan^2\beta} \quad (12)$$

While Equation (11) is for electric dipoles and Equation (12) is for electric rotors, showing an inverse expression compared with the magnetic ones. Due to the limited spectral resolution or the low emission intensities of the electric dipole and rotor peaks (**Figure S8,** Supporting Information), we can only resolve with sufficient clarity one of the electric dipole peaks (**Figure S9,** Supporting Information), obtaining the averaged polar angle $\alpha=60.1°$.

We further inspect the relation between the polar angles of dipoles (or rotors) and the crystal lattice structure. The lattice constants ($a = b = 5.96$ Å, $c = 3.53$ Å) of the host hexagonal microcrystal yields $arctan(c/a) = 30.6°$ and $arctan(a/c) = 59.4°$, which are around the same values of the dipole (or rotor) polar angles $\alpha$ and $\beta$ shown in Table 1, 2 and Table S2 (Supporting Information). Indeed, $\alpha$ (30.2°) and $\beta_I$ (59.4°) of the magnetic dipoles and rotors and $\alpha$ (60.1°) of the electric dipoles are of extremely high precision (<1°) of agreements, with the values of each sample shown in **Figure 4**. This correspondence can be understood as follows. The crystalline lattice serves an electrostatic perturbation on the electron orbitals of the doped $Eu^{3+}$ ions, defining a preferential direction c/a, as shown in the top-right inset of Figure 4. The magnetic rotor and electric dipole are oriented along c/a, while the magnetic dipole and electric rotor are oriented perpendicular to c/a. Although on each lattice site, the vector of a actually shows three directions with 120° between each other, the dipoles and rotors on each site have anyway to choose one direction of c/a out of the three, which, nevertheless, still leads to a macroscopic distribution of high order due to the overall hexagonal rotation symmetry, as already illustrated in Figure 3d. It is the incoherence between the individual dopant sites that leads to the partly linear polarization of emissions. Otherwise, the coherent superposition of all the moments of dipoles and rotors would leads to purely linearly polarized emissions, which will probably happen in the case of superradiance or stimulated emission. We note that the causal link between the lattice symmetry (i.e., the vector of c/a) and the LDOP may be essential for any type of crystals doped with any type of rare-earth ions.

In addition to the discovery of the physical association between the polarization of emission and the lattice structure, the precise polarization-resolved measurements provide an efficiently purely optical technology to identify the three-dimensional (3D) orientation of the single micro or nanocrystal. To demonstrate this practical application, we suppose the rod-shaped microcrystal is put in a 3D space described by a *x'-y'-z'* coordinate illustrated in **Figure 5**, which is then inspected with a top-down view, i.e., PL detection along z' direction. The orientation of the rod is than described by the polar and azimuthal angles ($\theta'$, $\varphi'$) in Figure 5. We define $I_{z'x'}$ and $I_{z'y'}$ as the light intensity of any of the magnetic or electric transition peaks detected along *z'* and polarized along *x'* and *y'*, respectively. From the geometry we obtain:

$$I_{z'x'} = I_\parallel \sin^2\theta' \sin^2\varphi' + I_\perp \cos^2\theta' + I_{ax} \sin^2\theta' \cos^2\varphi' \tag{13}$$

$$I_{z'y'} = I_\parallel \cos^2\theta' + I_\perp \sin^2\theta' \sin^2\varphi' + I_{ax} \sin^2\theta' \cos^2\varphi' \tag{14}$$

where $I_\parallel$ and $I_\perp$ are the measured intensities polarized parallel and perpendicular to the crystal c-axis, and the $I_{ax}$ is the expected intensity when measuring along the crystal c-axis with arbitrary polarization ($I_{ax}$ is not polarization-dependent as discussed in the Supporting Information). The orientation (θ′, φ′) can be derived combining the measured values of $I_{z'x'}$ and $I_{z'y'}$ of two magnetic dipole (or rotor) transition peaks with different LDOPs (see details in the Supporting Information):

$$\sin^2\theta' = \frac{\left(\frac{I_{2z'y'}}{I_{2z'x'}} - B'\right)(1-A') - \left(\frac{I_{1z'y'}}{I_{1z'x'}} - A'\right)(1-B')}{\left(\frac{I_{2z'y'}}{I_{2z'x'}} - \frac{I_{1z'y'}}{I_{1z'x'}}\right)(A'-1)(B'-1)} \tag{15}$$

$$\begin{cases} \sin^2\varphi' = \dfrac{\frac{I_{1z'y'}}{I_{1z'x'}}[\sin^2\theta'(A'-1)+1] - A'}{\sin^2\theta'(1-A')} \\ \sin^2\varphi' = \dfrac{\frac{I_{2z'y'}}{I_{2z'x'}}[\sin^2\theta'(B'-1)+1] - B'}{\sin^2\theta'(1-B')} \end{cases} \tag{16}$$

where the subscript 1 and 2 refer to the two chosen magnetic transition peaks, and A' and B' are the ratios $I_{1\parallel}/I_{1\perp}$ and $I_{2\parallel}/I_{2\perp}$. Herein we address that A' and B' are common properties of all microcrystals of the same type which is determined by the crystal structure. Therefore, in the measurement for determining the 3D orientation of the single micro or nanocrystal, $I_{z'x'}$ and $I_{z'y'}$ are the only items to measure, whilst $I_\parallel/I_\perp$ are regarded as known values obtained from measurements on other in-plane samples of the same type, ensuring the practical use of the technology. In addition, (θ′, φ′) can also be derived from the measurements on two electric dipole (or rotor) transition peaks with different LDOPs, or on one magnetic and one electric dipole (or rotor) transition peaks regardless of the LDOPs (see details in the Supporting Information).

## 4. Conclusion

In conclusion, we perform polarization-resolved spectroscopy on single hexagonal NaYF$_4$:Eu$^{3+}$ microcrystals under the non-resonance excitation, and reveal the causal link between the linear polarization degrees of the emission and the geometric parameters of the crystal lattice via quantitative demonstrations of unprecedented accuracy, based on which we develop a purely optical technology for determining the 3D orientation of a single micro or nanocrystal. Our results are essential for designing ion-doped crystals with on-demand emission of polarization properties, and would allow the instantaneous control of polarization via superradiance or stimulated emission by coupling the single micro or nanocrystals with optical cavities or metastructures. The technology of optically detected 3D orientation of nanocrystals is of potential interest for microscale sensing and imaging beyond diffraction limit in dynamical systems such as microfluids and living cells.

## 5. Methods

*Sample preparation and characterization*: Hexagonal $NaYF_4$ microcrystals doped with 5% $Eu^{3+}$ ions were synthesized by using hydrothermal method. First, 910 mg of $Y(NO_3)_3 \cdot 6H_2O$, 56 mg of $Eu(NO_3)_3 \cdot 6H_2O$ and 465 mg of ethylenediamine tetraacetic acid (EDTA) were dissolved in 20 mL of distilled water under stirring. Then, 1.3 g of NaF dissolved in 30 mL of distilled water was added dropwise to form the mixture. After 1 h continuous stirring, the mixture was transferred into a Teflon-lined stainless steel autoclave and heated at 180 °C for 24 h. The resulting precipitates were washed with distilled water and absolute ethanol by centrifugation in several times, dried at 80 °C for 12 h, and collected as white powders. The crystal structure of the powders was identified by X-ray powder diffraction analysis using Rigaku D/Max2550 with Cu K$α$ ($λ$ = 1.5406 Å) radiation under 40 kV, 50 mA. The scanning rate was 8°/min in the two-theta ranging from 10° to 80°. For the optical experiments, the as-prepared powders were suspended in ethanol solution and dispersed onto quartz substrates, forming a randomly in-plane distribution of single microcrystals.

*Single particle measurements*: The femtosecond pulsed Ti:Sapphire laser ($λ_{center}$ = 790 nm) is sent through a half-wave plate (10RP02-46, Newport) and a Glan-laser Calcite polarizer (10GL08AR.16, Newport) to produce a purely linearly polarized beam with tunable power. The beam is then focused on a frequency doubler with BBO nonlinear crystal, yielding a 395 nm excitation source matching with the $^7F_0 \rightarrow ^5L_6$ transition of the doped $Eu^{3+}$ ions. The excitation beam passes a harmonic beamsplitter (10Q20UF.HB1, Newport) to filter out the residue of 790 nm light, and its polarization is converted into circularity via a combination of a half-wave and a quarter-wave plate (10RP02-48 and 10RP04-48, Newport). In fact, the polarization of the excitation beam does not affect the polarization of the measured luminescence in our experiments, thanks to the non-resonance excitation scheme. The excitation beam then passes through a 50:50 nonpolarizing beamsplitter and focus on the targeted single microcrystal whose axis of length (crystalline c-axis) is oriented parallel to the surface of the quartz substrate. The substrate can be motivated and scanned in x-y-z directions by a nanopositioner connected with a piezoelectric controller. This enables the precise localization and detection of single microcrystals. The mixed lights containing the emitted luminescence and the scattered excitation beam from the microcrystal are collected by the same focusing objective (NA = 0.8, LMPLFLN100x, Olympus) and is further filtered by a 405 nm long-pass edge filter (BLP01-405R-25, Semrock) to transmit only the microcrystal luminescence. The transmitted luminescence is then passed through an achromatic half-wave or quarter-wave plate (10RP52-1B or 10RP54-1B, Newport) placed in a motorized precision rotation stage (KPRM1E, Thorlabs), followed by a visible wire grid linear polarizer (WP25M-VIS, Thorlabs) to a spectrometer (Shamrock SR-750-A, Andor) equipped with an EMCCD camera (DU970P-BVF, Andor). In the experiments, in order to eliminate the polarization preference of the spectrometer, we keep the linear polarizer oriented parallel to the entrance slit of the spectrometer (defined as 0°) and only rotate the half-wave or quarter-wave plate for polarization analysis. Under the Poincaré sphere (PS) analysis, we record the emission intensities of $I(0°)$, $I(90°)$, $I(45°)$, and $I(-45°)$ by rotating the axis of the half-wave plate at 0°, 45°, 22.5° and -22.5°, and of $I(\sigma^+)$ and $I(\sigma^-)$ by rotating the axis of the quarter-wave plate at 45° and -45°, respectively. When using the polarization fitting (PF) analysis, we record a series of the emission intensity of $I(\theta)$ from $\theta = 0°$ to $\theta = 360°$ with a step of 15° by rotating the axis of the half-wave plate from 0° to 180° with a step of 7.5°. All emission intensities are taken with the same exposure time of 10 s.

**Acknowledgements**

This work is supported by National Natural Science Foundation of China (61890961, 12074303, 11804267 and 11904279) and Shaanxi Key Science and Technology Innovation Team Project (2021TD-56).

**Author Contributions**

P.L., Y.X.G. and F.L. conceived the idea and designed the project. F.L. supervised the project. P.L., Y.X.G., A.L., X.Y. and Y.P.Z. conducted the experiments and analyzed the data. P.L. and A.L. set up the theoretical model. Y.X.G., T.L.Y. and J.P.Z. contributed to sample preparations and X-ray diffraction measurements. P.L. and F.L. wrote the paper with inputs from all authors.

**Conflict of Interest**

The authors declare no competing financial interest.

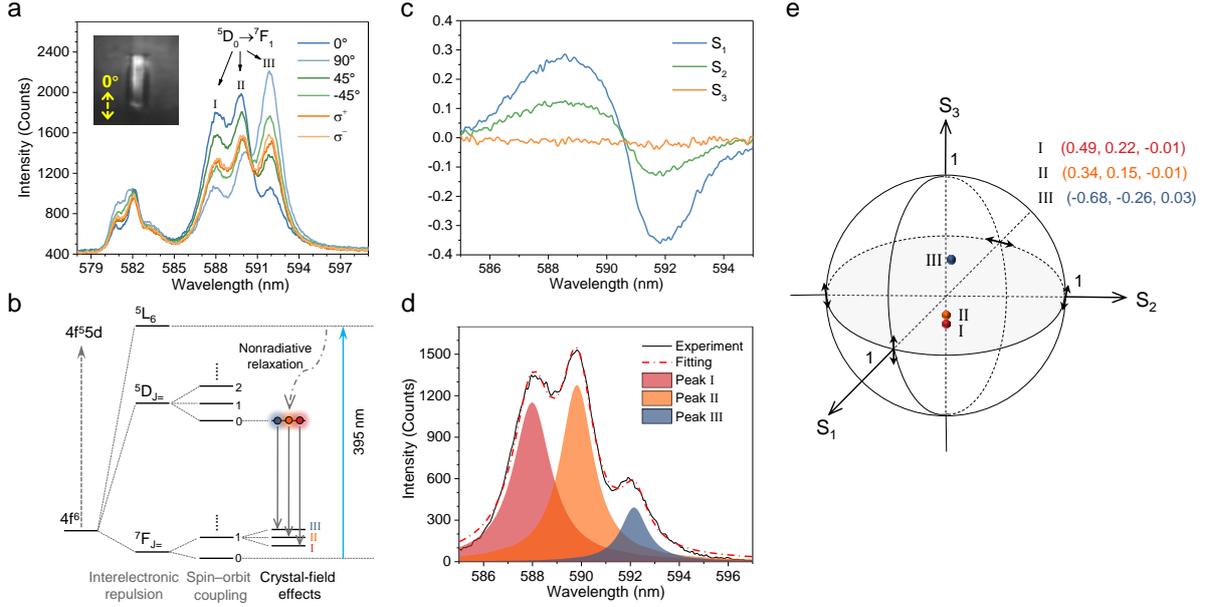

**Figure 1.** Poincaré sphere analysis of polarized photoluminescence of a single hexagonal NaYF$_4$:Eu$^{3+}$ microcrystal labeled as sample No. 1. **a)** Photoluminescence (PL) spectra of Eu$^{3+}$ ions doped in a single NaYF$_4$ microcrystal recorded at the six Stokes basis under an excitation of 395 nm, respectively. The three well-resolved peaks marked by I, II and III come from the $^5D_0 \rightarrow {}^7F_1$ magnetic transition. The inset is a CCD image showing the microcrystal lying on a substrate plane whose axis of length (crystalline c-axis) is oriented with a small angle to the spectrometer entrance slit which is defined as 0°. **b)** Schematic illustration of partial energy levels (4f$^6$) of Eu$^{3+}$ dopants in the microcrystal. The degeneracy of the $^7F_1$ level is totally lifted to three sublevels by the crystal-field perturbations. The downward solid arrows indicate the occurrence of luminescence from the $^5D_0$ excited state to the $^7F_1$ sublevels after the excitation at 395 nm. **c)** The three Stokes parameters calculated directly from the PL spectra shown in (**a**) using Equation (1-3) in the range of 585-595 nm corresponding to the $^5D_0 \rightarrow {}^7F_1$ transition. **d)** Spectral fitting analysis in the $^5D_0 \rightarrow {}^7F_1$ transition range by using the PL spectrum of 0° polarization from (**a**) as an example. The spectrum is well fitted by the sum of three Lorentzian peaks in which baseline is taken as a flat line with averaged intensity between 600 nm and 605 nm where Eu$^{3+}$ luminescence is not present. The peaks with their respective area integrals are highlighted in different colors. **e)** Three sets of Stokes parameters ($S_1$, $S_2$, $S_3$) corresponding to the three emission peaks of the $^5D_0 \rightarrow {}^7F_1$ transition mapped in a normalized Poincaré sphere with different locations. The Stokes parameters are calculated from the integrated intensities of the peaks using Equation (1-3). The Stokes parameters of all optical transitions locate on the equator plane indicating zero circular polarization degree.

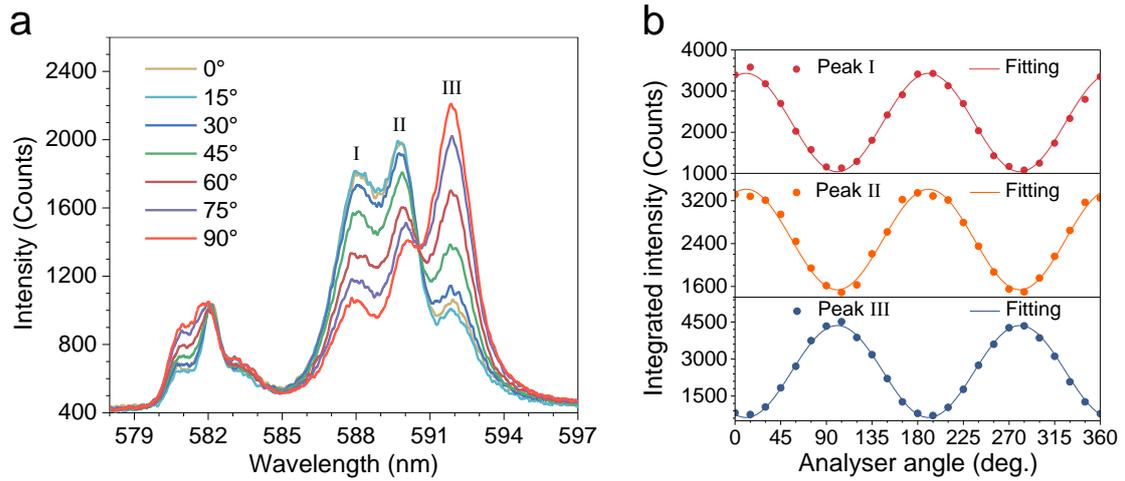

**Figure 2.** Polarization fitting analysis of polarized photoluminescence of the single NaYF$_4$:Eu$^{3+}$ microcrystal. **a**) Polarized PL spectra of the same single NaYF$_4$:Eu$^{3+}$ microcrystal with varied analyzer angles with a step of 15°. **b**) Polarization fittings of the three emission peaks of the $^5D_0\rightarrow{}^7F_1$ transition based on their Lorentzian integrated intensities at the varied analyzer angles using Equation (4-2).

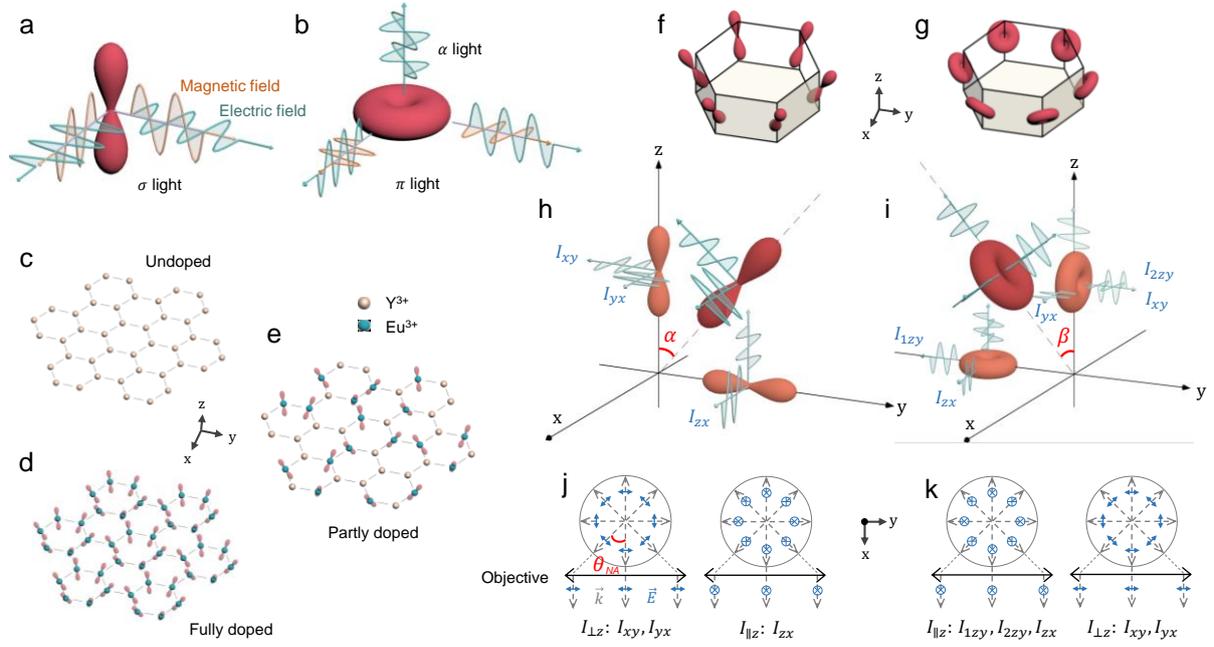

**Figure 3.** Orientation analysis of the magnetic transition dipoles and rotors of $Eu^{3+}$ ions in the $NaYF_4$ microcrystals. **a,b)** Schematic illustrations of radiations by electromagnetic fields in the magnetic dipole (**a**) and rotor (**b**). The dipole radiates the $\sigma$ light, whereas the rotor emits the $\pi$ and $\alpha$ lights, simultaneously. **c-e)** Illustrations of $Y^{3+}$ sites (**c**) and the sites fully (**d**) or partly (**e**) doped by $Eu^{3+}$ ions with their magnetic dipoles distributed in the hexagonal lattices of the $NaYF_4$ microcrystal whose c-axis being parallel to the z-axis. **f,g)** Effective macroscopic distributions of the ensembles of the dipoles (**f**) and rotors (**g**) according to the crystal hexagonal rotational symmetry. **h,i)** Schematic illustrations of the emissions from a single effective dipole (**h**) and a single rotor (**i**) having a polar angle $\alpha$ or $\beta$ with respect to the z-axis (c-axis), respectively. **j,k)** Illustrations of the light path emitted by all effective dipoles (**j**) and rotors (**k**) which are collected by the objective lens with a numerical aperture of $\theta_{NA}$, resulting in the light intensities parallel $I_\parallel$ and perpendicular $I_\perp$ to the c-axis. The blue arrows (or crosses) represent the direction of polarization while the grey arrows represent the direction of propagation.

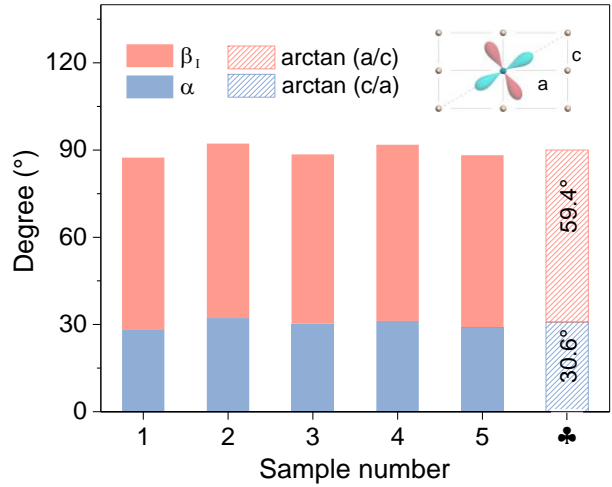

**Figure 4.** Relations between the orientations of transition dipoles and rotors and the crystal lattice symmetry. The red and blue columns represent the polar angles of the magnetic rotors for the transition peak I and of the magnetic dipoles for the transition peak III, respectively, based on the polarization fitting analysis from the measured five samples. The values are then compared with $arctan\,(a/c) = 59.4°$ and $arctan\,(c/a) = 30.6°$ calculated from the host crystal lattice constants. The inset is a schematic illustration of the orientations of electric (blue) and magnetic (red) dipoles in the hexagonal $NaYF_4$ crystal lattice as viewed along the x-axis.

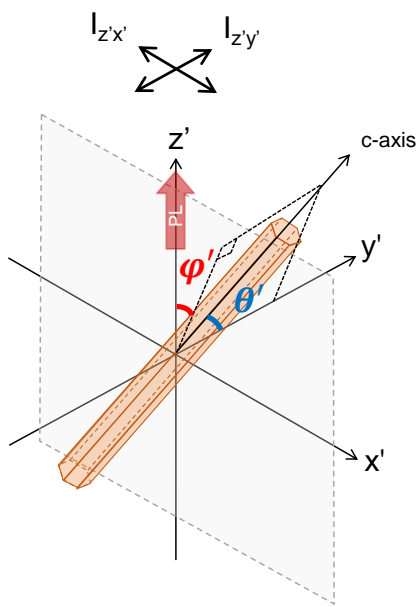

**Figure 5.** Three-dimensional (3D) orientation analysis of a single hexagonal NaYF$_4$:Eu$^{3+}$ microcrystal. A rod-shaped single microcrystal is placed in a *x'-y'-z'* coordinate, with its 3D orientation described by the polar and azimuthal angles ($\theta'$, $\varphi'$). The emission intensities collected along z' with two orthogonal polarization angles are labeled as $I_{z'x'}$ and $I_{z'y'}$, respectively.

**Table 1.** Polarization and the corresponding magnetic dipole or rotor orientations of the $^5D_0 \rightarrow {}^7F_1$ transition peaks from the Poincaré sphere analysis. Linear polarization angle $\varphi$ and degree (LDOP) for the three emission peaks of the $^5D_0 \rightarrow {}^7F_1$ transition are calculated from their respective ($S_1$, $S_2$, $S_3$) parameters, and the polar angle $\alpha$ or $\beta$ of the corresponding magnetic dipoles or rotors are deduced from the LDOP using Equation (8) and (10) among the five measured microcrystals, respectively. The average values of the LDOP, $\alpha$ and $\beta$ are indicated in the last row.

| Sample No. | Poincaré sphere analysis | | | | | | | | |
|---|---|---|---|---|---|---|---|---|---|
| | Peak I | | | Peak II | | | Peak III | | |
| | $\varphi$ | LDOP | $\beta_I$ | $\varphi$ | LDOP | $\beta_{II}$ | $\varphi$ | LDOP | $\alpha$ |
| 1 | 11.9° | 0.54 | 59.3° | 12.0° | 0.38 | 52.6° | 100.4° | 0.73 | 29.2° |
| 2 | 95.7° | 0.56 | 60.1° | 96.1° | 0.39 | 53.0° | 8.9° | 0.67 | 32.4° |
| 3 | 8.8° | 0.51 | 58.1° | 8.4° | 0.35 | 51.5° | 98.7° | 0.70 | 31.0° |
| 4 | 104.4° | 0.56 | 60.2° | 105.3° | 0.32 | 50.3° | 13.9° | 0.63 | 33.9° |
| 5 | 10.1° | 0.54 | 59.5° | 9.4° | 0.40 | 53.5° | 99.0° | 0.73 | 29.2° |
| Mean | | 0.54 | 59.4° | | 0.37 | 52.2° | | 0.69 | 31.1° |

**Table 2.** Polarization and the corresponding magnetic dipole or rotor orientations of the $^5D_0 \rightarrow {}^7F_1$ transition peaks from the polarization fitting analysis. Intensities of the orthogonal components (A, B) and linear polarization angle $\phi$ are obtained from the fittings of the three emission peaks of the $^5D_0 \rightarrow {}^7F_1$ transition using Equation (4-2) for the five measured microcrystals. Linear polarization degree (LDOP) of the three emission peaks are calculated from A and B, and the polar angle $\alpha$ or $\beta$ of the corresponding magnetic dipoles or rotors are deduced from the LDOP using Equation (8) and (10), respectively. The average values of the LDOP, $\alpha$ and $\beta$ are indicated in the last row.

| Sample No. | Polarization fitting analysis ||||||||||||||| 
| | Peak I ||||| Peak II ||||| Peak III ||||| 
| | A | B | $\phi$ | LDOP | $\beta_I$ | A | B | $\phi$ | LDOP | $\beta_{II}$ | A | B | $\phi$ | LDOP | $\alpha$ |
|---|---|---|---|---|---|---|---|---|---|---|---|---|---|---|---|
| 1 | 3432 | 1039 | 10.8° | 0.54 | 59.1° | 3413 | 1540 | 10.8° | 0.38 | 52.6° | 4357 | 630 | 100.8° | 0.75 | 28.3° |
| 2 | 2553 | 728 | 97.7° | 0.56 | 60.0° | 2236 | 1079 | 97.7° | 0.35 | 51.4° | 3154 | 626 | 7.7° | 0.67 | 32.2° |
| 3 | 2517 | 804 | 8.0° | 0.52 | 58.3° | 2387 | 1131 | 8.0° | 0.36 | 51.8° | 3173 | 539 | 98.0° | 0.71 | 30.2° |
| 4 | 2995 | 820 | 105.3° | 0.57 | 60.6° | 2722 | 1329 | 105.3° | 0.34 | 51.2° | 3908 | 717 | 15.3° | 0.69 | 31.2° |
| 5 | 4146 | 1248 | 9.5° | 0.54 | 59.2° | 3840 | 1835 | 9.5° | 0.35 | 51.6° | 5158 | 794 | 99.5° | 0.73 | 29.0° |
| Mean | | | | 0.55 | 59.4° | | | | 0.36 | 51.7° | | | | 0.71 | 30.2° |

Supporting Information

**Deterministic relation between optical polarization and lattice symmetry revealed in ion-doped single microcrystals**

*Peng Li,*[1,†,*] *Yaxin Guo,*[1,†] *Ao Liu,*[1] *Xin Yue,*[1] *Taoli Yuan,*[2] *Jingping Zhu,*[1] *Yanpeng Zhang,*[1] *and Feng Li*[1,*]

[1]Key Laboratory for Physical Electronics and Devices of the Ministry of Education, School of Electronic Science and Engineering, Faculty of Electronic and Information Engineering, Xi'an Jiaotong University, Xi'an 710049, P. R. China

[2]School of Electronic Information and Artificial Intelligence, Shaanxi University of Science and Technology, Xi'an 710021, P. R. China

†These authors contributed equally
*Corresponding author: ponylee@stu.xjtu.edu.cn; felix831204@xjtu.edu.cn

# I. Analysis of the partially linearly polarized optical transitions via the Poincaré sphere and the polarization fitting methods (Figure S1-S9, Tables S1 and S2)

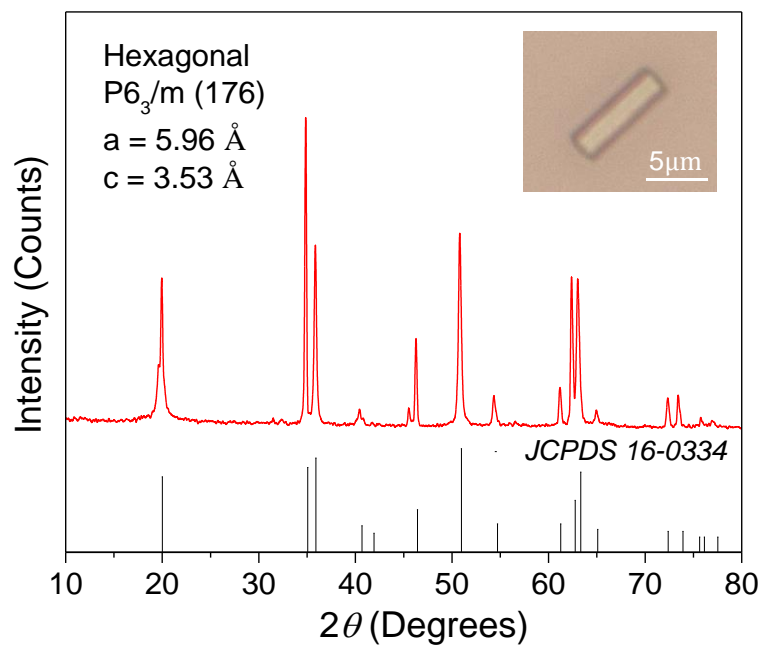

**Figure S1.** X-ray diffraction pattern (red line) of the as-synthesized NaYF$_4$:Eu$^{3+}$ microcrystals with the standard hexagonal reference card (JCPDS card *no. 16-0334*). The microcrystals crystallize in P6$_3$/m space group with a hexagonal structure ($a = b = 5.96$ Å, $c = 3.53$ Å, $\alpha = \beta = 90^o, \gamma = 120^o$). The inset is an optical microscope image of a single one from the as-synthesized microcrystals randomly dispersed on a quartz substrate, showing a dimension ≈ 8 μm × 2 μm.

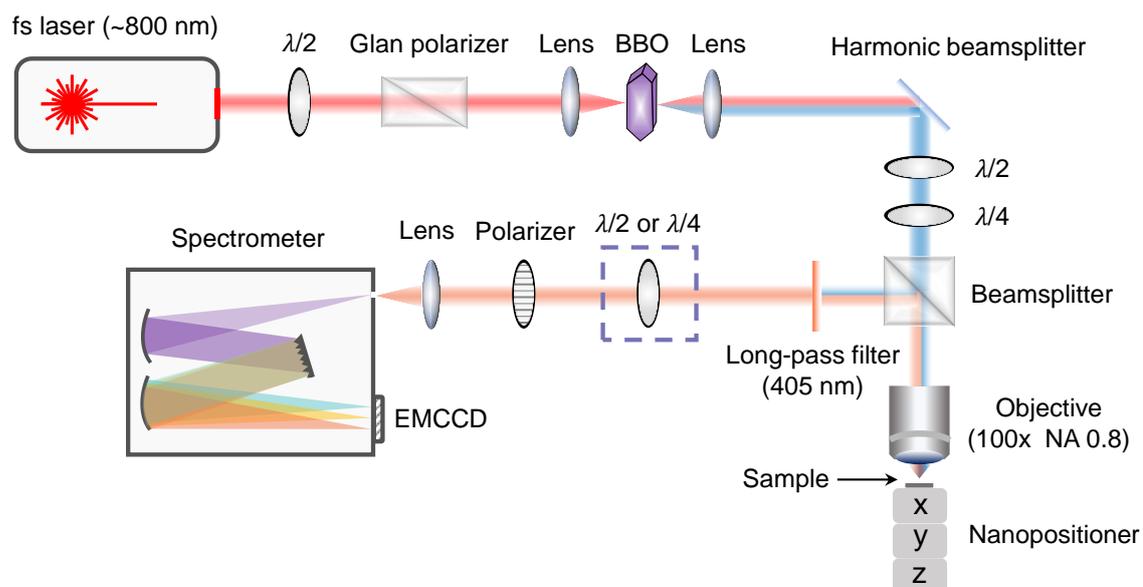

**Figure S2.** A scheme of the experimental setup for polarization-resolved spectroscopy of the in-plane single microcrystals.

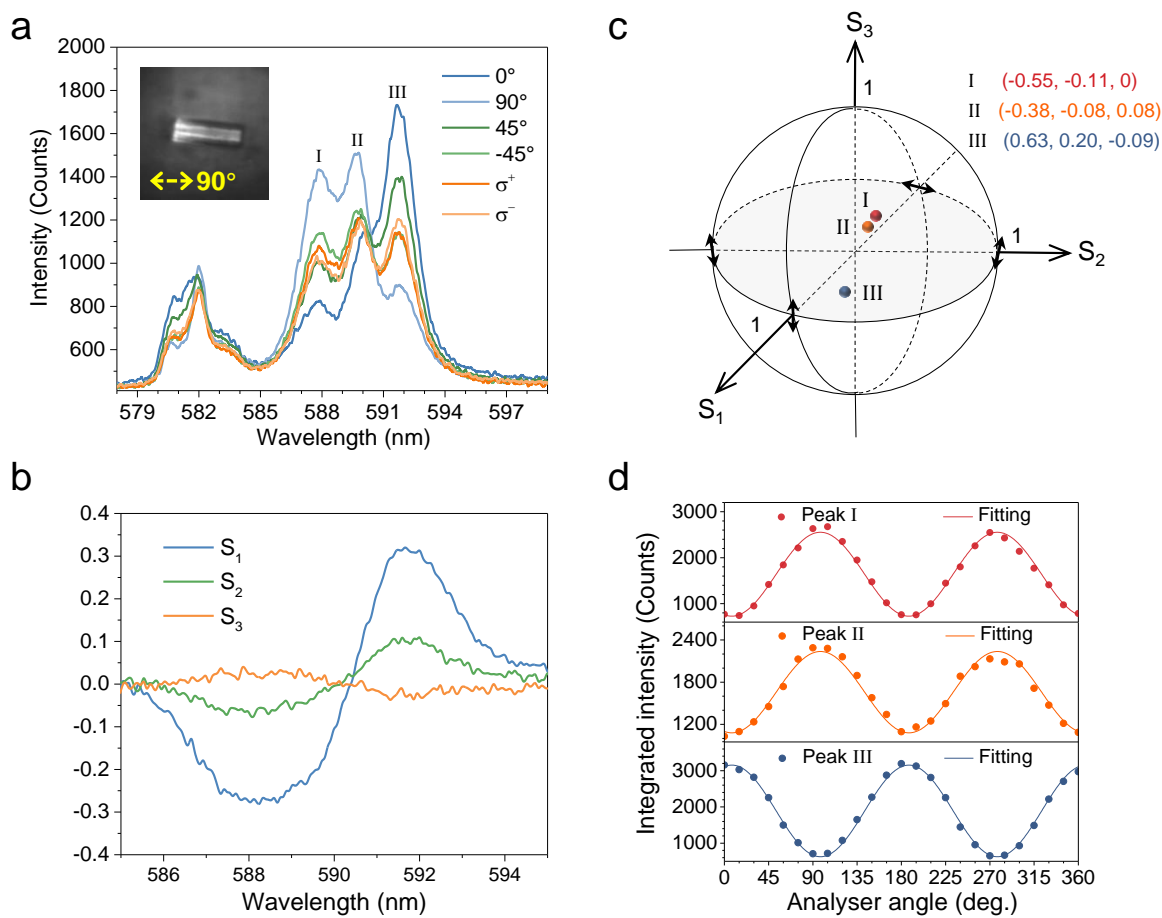

**Figure S3.** Poincaré sphere analysis and polarization fitting analysis of polarized photoluminescence of a single microcrystal labeled as sample No. 2. **a)** PL spectra of a single NaYF$_4$:Eu$^{3+}$ microcrystal recorded at the six Stokes basis under a 395 nm excitation, respectively. The inset is a CCD image showing the microcrystal oriented with a small angle perpendicular to the spectrometer entrance slit. **b)** The S$_1$, S$_2$ and S$_3$ calculated directly from the PL spectra shown in (**a**) in the range of 585-595 nm corresponding to the $^5D_0 \rightarrow {}^7F_1$ transition. **c)** Three sets of Stokes parameters (S$_1$, S$_2$, S$_3$) corresponding to the three emission peaks of the $^5D_0 \rightarrow {}^7F_1$ transition mapped in a normalized Poincaré sphere, respectively. The Stokes parameters are calculated from the Lorentzian integrated intensities of the peaks. **d)** Polarization fittings of the three peaks using the integrated intensities at the varied analyzer angles, respectively.

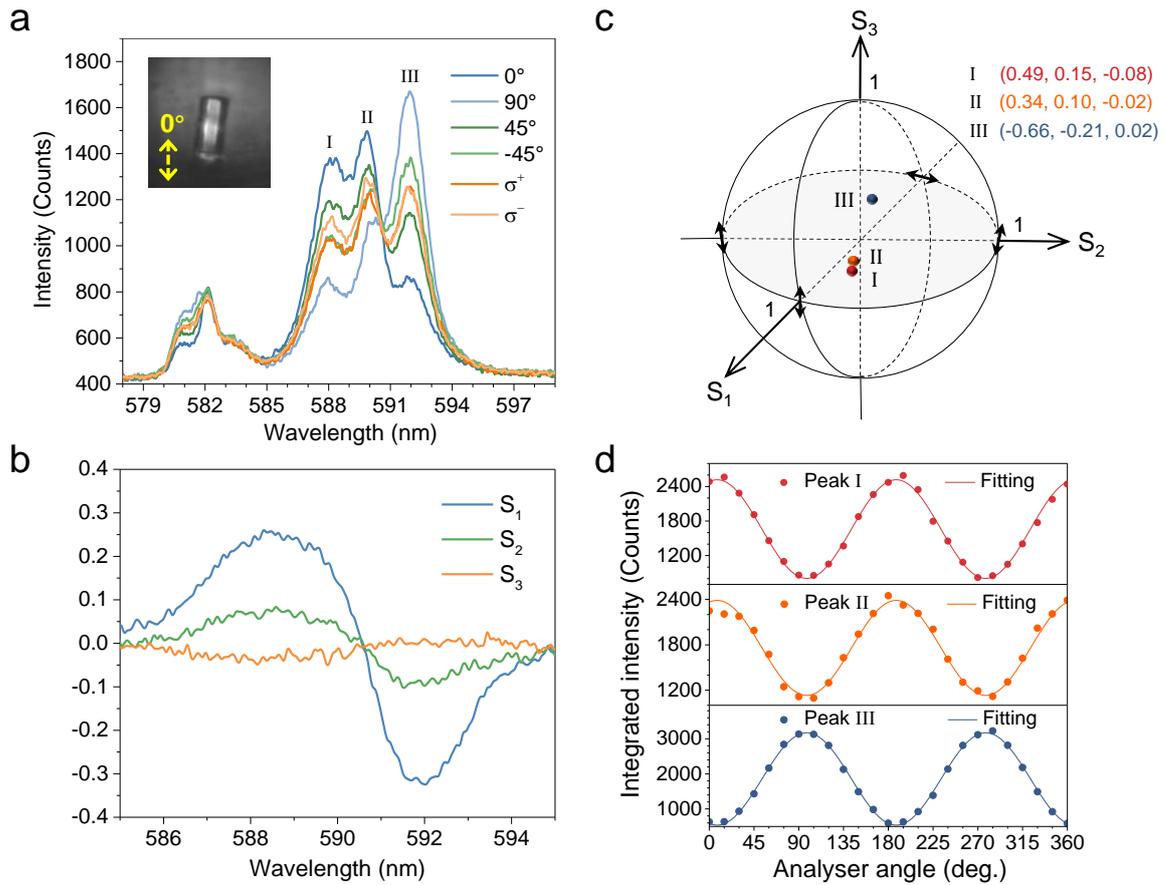

**Figure S4.** Poincaré sphere analysis and polarization fitting analysis of polarized photoluminescence of a single microcrystal labeled as sample No. 3. **a)** PL spectra of a single NaYF$_4$:Eu$^{3+}$ microcrystal recorded at the six Stokes basis under a 395 nm excitation, respectively. The inset is a CCD image showing the microcrystal oriented with a small angle to the spectrometer entrance slit. **b)** The $S_1$, $S_2$ and $S_3$ calculated directly from the PL spectra shown in (**a**) in the range of 585-595 nm corresponding to the $^5D_0 \rightarrow {}^7F_1$ transition. **c)** Three sets of Stokes parameters ($S_1$, $S_2$, $S_3$) corresponding to the three emission peaks of the $^5D_0 \rightarrow {}^7F_1$ transition mapped in a normalized Poincaré sphere, respectively. The Stokes parameters are calculated from the Lorentzian integrated intensities of the peaks. **d)** Polarization fittings of the three peaks using the integrated intensities at the varied analyzer angles, respectively.

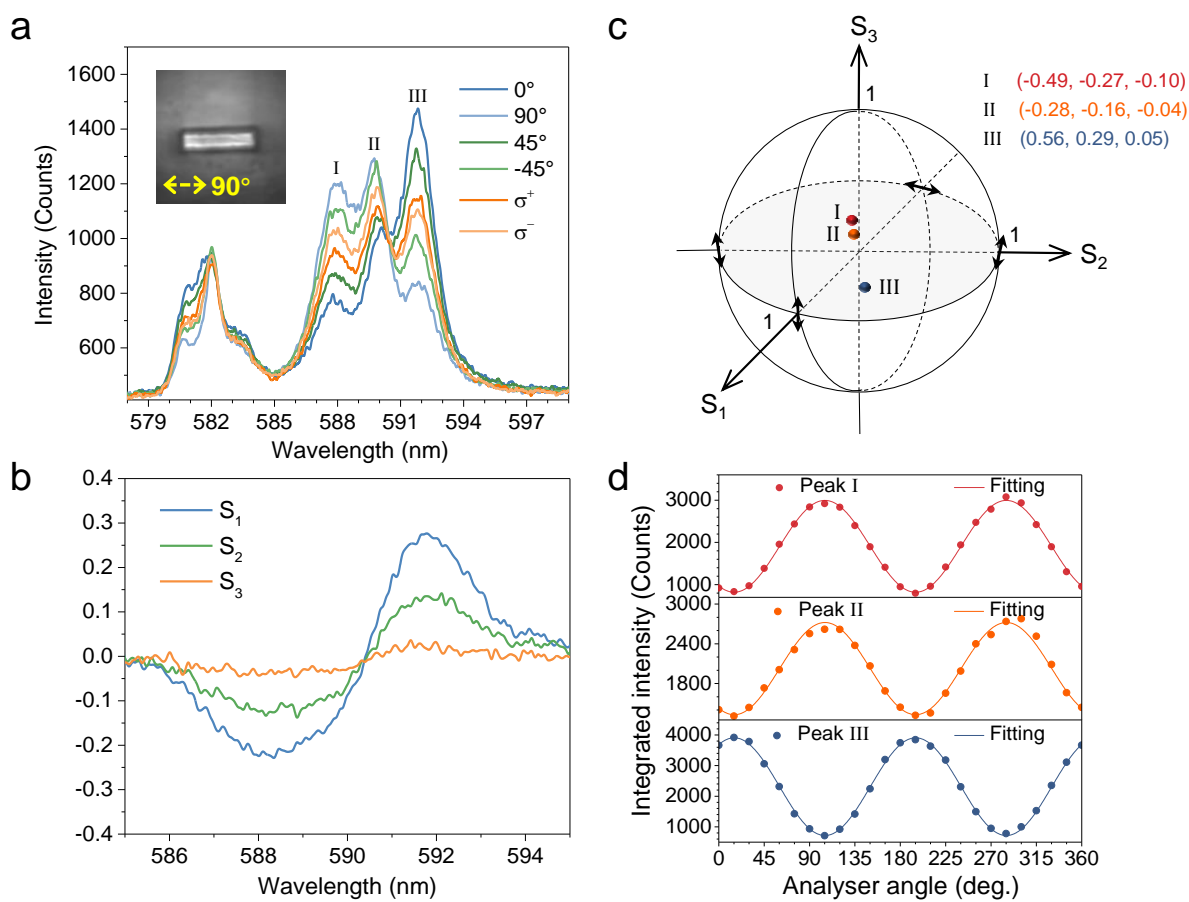

**Figure S5.** Poincaré sphere analysis and polarization fitting analysis of polarized photoluminescence of a single microcrystal labeled as sample No. 4. **a)** PL spectra of a single $NaYF_4:Eu^{3+}$ microcrystal recorded at the six Stokes basis under a 395 nm excitation, respectively. The inset is a CCD image showing the microcrystal oriented with a small angle perpendicular to the spectrometer entrance slit. **b)** The $S_1$, $S_2$ and $S_3$ calculated directly from the PL spectra shown in (**a**) in the range of 585-595 nm corresponding to the $^5D_0 \rightarrow {}^7F_1$ transition. **c)** Three sets of Stokes parameters ($S_1$, $S_2$, $S_3$) corresponding to the three emission peaks of the $^5D_0 \rightarrow {}^7F_1$ transition mapped in a normalized Poincaré sphere, respectively. The Stokes parameters are calculated from the Lorentzian integrated intensities of the peaks. **d)** Polarization fittings of the three peaks using the integrated intensities at the varied analyzer angles, respectively.

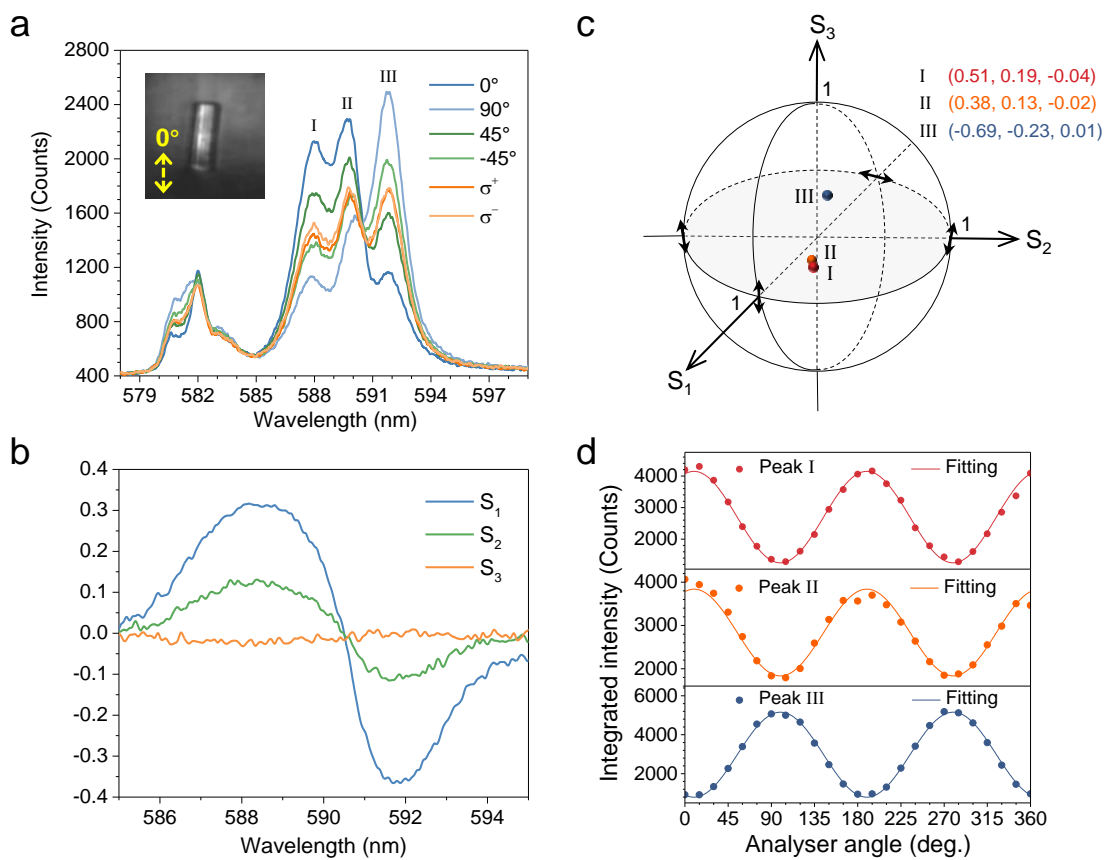

**Figure S6.** Poincaré sphere analysis and polarization fitting analysis of polarized photoluminescence of a single microcrystal labeled as sample No. 5. **a)** PL spectra of a single NaYF$_4$:Eu$^{3+}$ microcrystal recorded at the six Stokes basis under a 395 nm excitation, respectively. The inset is a CCD image showing the microcrystal oriented with a small angle to the spectrometer entrance slit. **b)** The S$_1$, S$_2$ and S$_3$ calculated directly from the PL spectra shown in (**a**) in the range of 585-595 nm corresponding to the $^5D_0 \rightarrow {}^7F_1$ transition. **c)** Three sets of Stokes parameters (S$_1$, S$_2$, S$_3$) corresponding to the three emission peaks of the $^5D_0 \rightarrow {}^7F_1$ transition mapped in a normalized Poincaré sphere, respectively. The Stokes parameters are calculated from the Lorentzian integrated intensities of the peaks. **d)** Polarization fittings of the three peaks using the integrated intensities at the varied analyzer angles, respectively.

**Table S1.** Irreducible representations of the initial ($^5D_0$) and terminating ($^7F_{J=0, 1, 2}$) states of $Eu^{3+}$ dopants in $C_s$ site symmetry group and the allowed oscillating forms or polarization directions (by the selection rules) for electromagnetic transitions between the initial and terminating representations.[1-3]

| Initial state | Irreducible representations[a] | Terminating state | Irreducible representations[a] | Transforms like | Number of peaks | Transition nature |
|---|---|---|---|---|---|---|
| $^5D_0$ | A'/$\Gamma_1$ | $^7F_0$ | A'/$\Gamma_1$ | (x, y) | 1 | ED |
| | | $^7F_1$ | A'/$\Gamma_1$ | $R_z$ | 1 | MD |
| | | | 2A''/2$\Gamma_2$ | ($R_x$, $R_y$) | 2 | |
| | | $^7F_2$ | 3A'/3$\Gamma_1$ | (x, y) | 3 | ED |
| | | | 2A''/2$\Gamma_2$ | z | 2 | |

[a] A' and A'' refer to Mulliken symbol; and $\Gamma_1$ and $\Gamma_2$ to Bethe symbol.

In the table, x, y, and z correspond to electric dipole (ED) operators oscillating linearly along x, y, and z; and $R_x$, $R_y$, and $R_z$ to magnetic dipole (MD) operators oscillating linearly along x, y, and z. The notation (x, y) stands for interchangeable x and y operators, which can be considered as an electric rotor oscillating in a x-y plane; and ($R_x$, $R_y$) for interchangeable $R_x$ and $R_y$ operators as a magnetic rotor oscillating in a x-y plane. The symbols x, y and z here are used to express the allowed transition forms, following the well-established expression rules. There is no relation between the symbols here and the x-y-z coordinates presented in **Figure 3** and **Figure S7**.

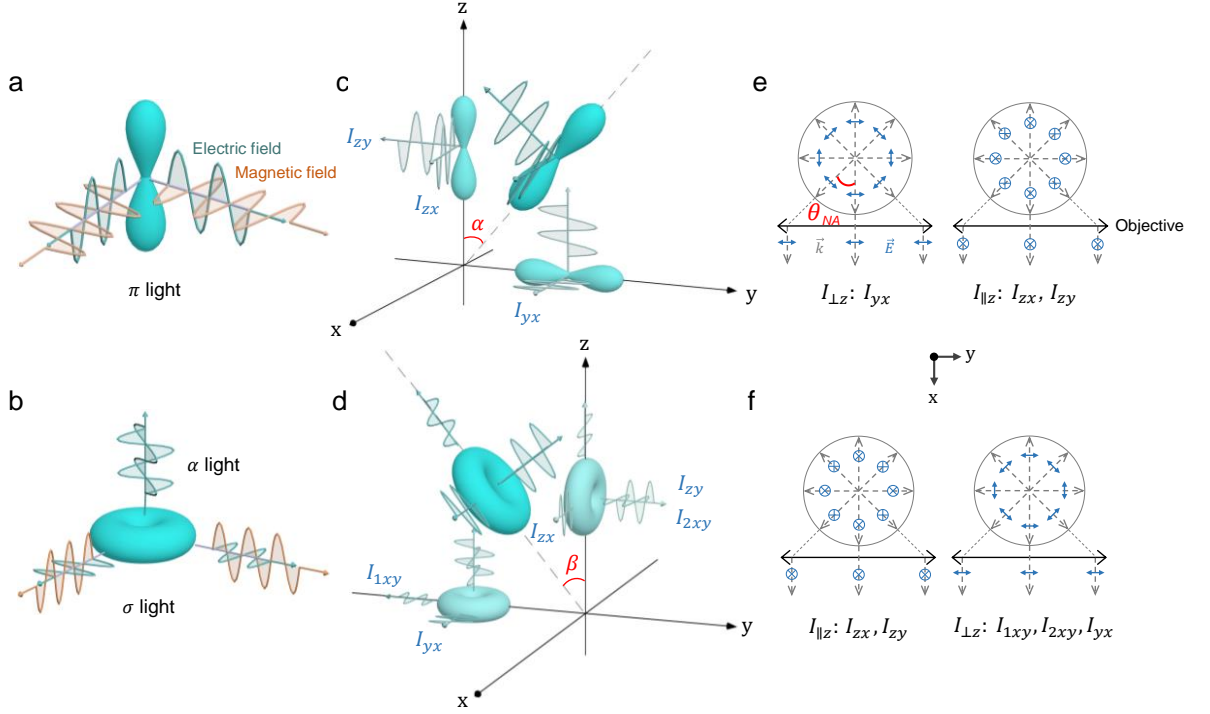

**Figure S7.** Orientation analysis of the electric transition dipoles and rotors of $Eu^{3+}$ ions in the NaYF$_4$ microcrystals. **a,b)** Schematic illustrations of radiations by electromagnetic fields in the electric dipole (**a**) and rotor (**b**). The dipole radiates the $\pi$ light, whereas the rotor emits the $\sigma$ and $\alpha$ lights, simultaneously. **c,d)** Schematic illustrations of the emissions from a single effective dipole (**c**) or rotor (**d**) having a polar angle $\alpha$ or $\beta$ with respect to the z-axis (crystalline c-axis), respectively. **e,f)** Illustrations of the light path emitted by all effective dipoles (**e**) and rotors (**f**) which are collected by the objective lens with a numerical aperture of $\theta_{NA}$, resulting in the light intensities parallel $I_\parallel$ and perpendicular $I_\perp$ to the c-axis. The blue arrows (or crosses) represent the direction of polarization while the grey arrows represent the direction of propagation. Herein we have $I_{yx} = I \cdot sin^2\alpha$ and $I_{zx} = I_{zy} = I \cdot cos^2\alpha$ for the dipoles, and of $I_{1xy} = I_{yx} = I \cdot sin^2\beta$ and $I_{zx} = I_{zy} = I_{2xy} = I \cdot cos^2\beta$ for the rotors, respectively.

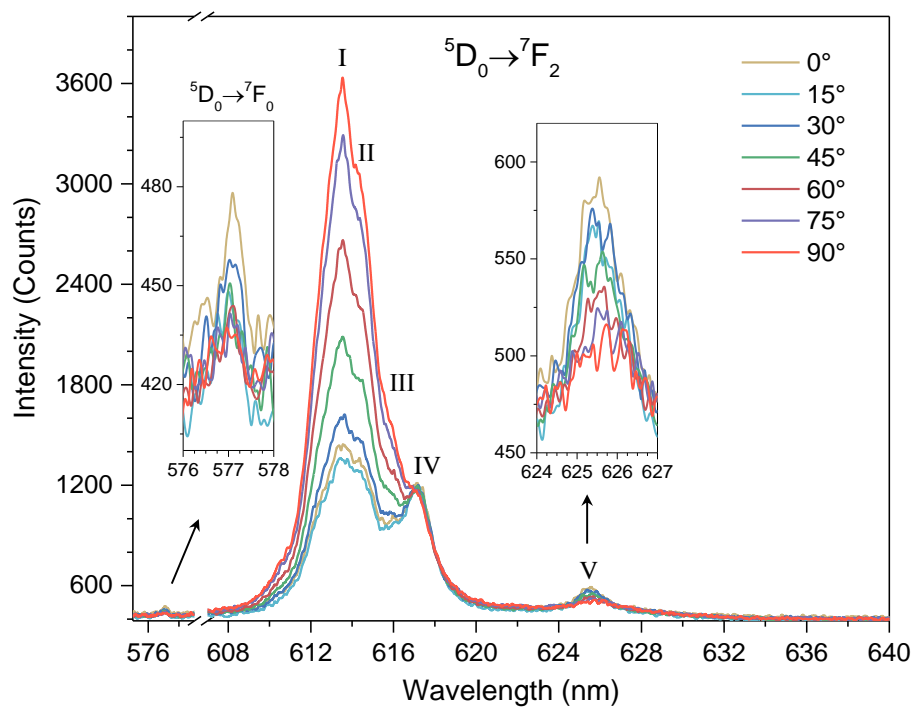

**Figure S8.** Polarized photoluminescence from the electric transitions in the single NaYF$_4$:Eu$^{3+}$ microcrystal (sample No. 1) with varied analyzer angles. Two emission bands of 576-578 nm and 608-632 nm correspond to $^5D_0\rightarrow{}^7F_0$ and $^5D_0\rightarrow{}^7F_2$ electric transitions of the Eu$^{3+}$ dopants, respectively. The $^5D_0\rightarrow{}^7F_0$ transition and the peaks I, II and III from the $^5D_0\rightarrow{}^7F_2$ transition are assigned to the electric rotor oscillating, and the peaks IV and V from the $^5D_0\rightarrow{}^7F_2$ transition are assigned to the electric dipole oscillating, respectively, according to the selection rules in **Table S1**. As the intensity of the $^5D_0\rightarrow{}^7F_0$ is too weak and undulatory and the peaks I-IV of the $^5D_0\rightarrow{}^7F_2$ transition are merged together, we analyze the peak V, as shown in **Figure S9**.

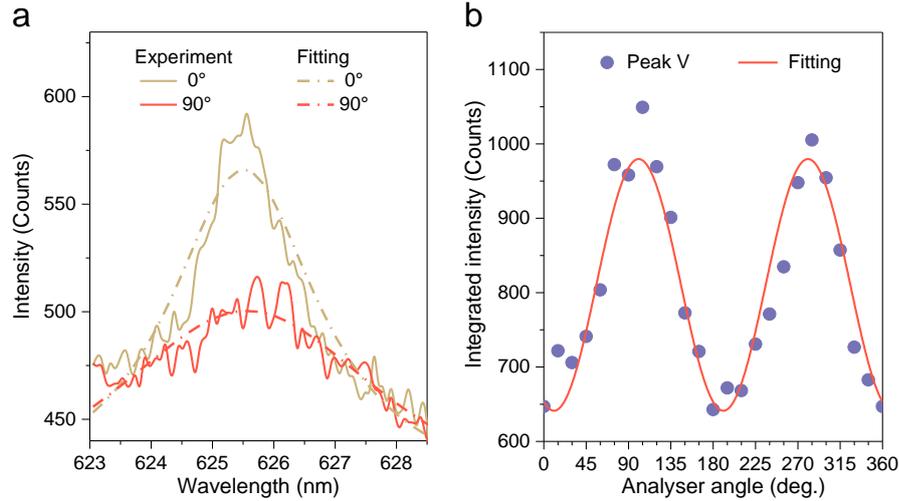

**Figure S9.** Polarization fitting analysis of the peak V from the electric $^5D_0 \to {}^7F_2$ transition. **a**) Spectral fitting analysis of the peak V by using 0° and 90° PL spectra taken from **Figure S8** as examples. The spectra are also fitted well by the Lorentzian peaks in which baseline is taken as a flat line with averaged intensity between 634 nm and 640 nm where $Eu^{3+}$ luminescence is not present. **b**) Polarization fittings of the peak V based on the integrated intensities at the varied analyzer angles using **Equation (4-2)** of the main text. Although at the peak point more counts are read at 0° than that at 90° in **(a)**, the integrated intensity of the whole peak area shows an opposite situation **(b)**.

**Table S2.** Polarization and the corresponding electric dipole orientation of the peak V from the polarization fitting analysis and the poincaré sphere analysis in the five measured single microcrystals, respectively. The average values of LDOP, $\alpha$ and $\beta$ are indicated in the last row.

| Sample No. | $^5D_0 \rightarrow {}^7F_2$: Peak V | | | | | | | | |
|---|---|---|---|---|---|---|---|---|---|
| | Polarization fitting analysis | | | | | Poincaré sphere analysis | | | |
| | A | B | $\phi$ | LDOP | $\alpha$ | $(S_1, S_2, S_3)$ | $\varphi$ | LDOP | $\alpha$ |
| 1 | 980 | 641 | 100.8° | 0.21 | 60.2° | (-0.18, -0.07, -0.04) | 100.9° | 0.20 | 59.9° |
| 2 | 730 | 463 | 7.7° | 0.22 | 60.6° | (0.21, 0.08, 0.03) | 9.9° | 0.22 | 60.6° |
| 3 | 804 | 517 | 98.0° | 0.22 | 60.4° | (-0.21, -0.14, -0.04) | 106.3° | 0.25 | 61.3° |
| 4 | 863 | 591 | 15.3° | 0.19 | 59.7° | (0.11, 0.06, -0.02) | 14.1° | 0.13 | 58.2° |
| 5 | 1147 | 737 | 90.5° | 0.22 | 60.4° | (-0.17, -0.07, -0.03) | 101.3° | 0.18 | 59.5° |
| Mean | | | | 0.21 | 60.3° | | | 0.20 | 59.9° |

## II. Determination of three-dimensional (3D) orientation of a single microcrystal via polarization-resolved spectroscopy

In the main text, we have the c-axis of single microcrystals lying on the plane of the substrate, and thereby are able to detect the two orthogonally polarized intensities $I_\parallel$ and $I_\perp$ of any transition peak, both propagating normal to the c-axis of the microcrystal which is our direction of detection. However, if the rod-shaped microcrystal shows an arbitrary 3D orientation, then the detection (which is supposed to be normal to the substrate) is no longer normal to the c-axis. This is clearly illustrated in **Figure S10**, where the substrate is the x'-y' plane and the detection is along the z' axis. The orientation of the rod is then described by the polar and azimuthal angles $(\theta', \varphi')$. It is obvious that the values of $(\theta', \varphi')$ cannot be derived just using $I_\parallel$ and $I_\perp$ as they are intensities corresponding to only 2D in-plane orientations. In order to obtain 3D orientation information, it seems an extra base for light propagating along the c-axis (noted as $I_{ax}$ hereafter) of the microcrystal is needed, which is technically very difficult to measure as it's quite impossible to put a rod standing upright on the substrate. Nevertheless, our analysis with the dipoles and rotors confirms that there exists relation between $I_\parallel$, $I_\perp$ and $I_{ax}$, making the problem much simply. We take the situation of **Figure 3h** in the main text as an example. The magnetic sub-dipole lying along y direction has an optical field propagating along z which constitutes the only contribution to $I_{ax}$, therefore $I_{ax} = I_{zx}$ for a single dipole. Then when the dipole rotates with the distribution in **Figure 3j**, $I_{ax} = I_\parallel$ holds for the integrated light as $I_{zx} = I_\parallel$ after the integration. Herein the $I_{ax}$ is unpolarized as its polarization angle rotates a round along z. The situation for the magnetic rotor is more complicated containing three components contributing to $I_{ax}$, but the same rule $I_{ax} = I_\parallel$ still holds. Therefore $I_{ax} = I_\parallel$ holds for all magnetic transition peaks. Similarly, we derive $I_{ax} = I_\perp$ for all electric transition peaks. These rules have also been demonstrated experimentally,[4] further confirming the robustness of our method as follows.

Considering a single hexagonal NaYF$_4$:Eu$^{3+}$ microcrystal with an arbitrary 3D orientation by polar and azimuthal angles $(\theta', \varphi')$ in a x'-y'-z' coordinate system (**Figure S10**), the emission intensities of any transition peaks collected at two orthogonal polarization angles, $I_{z'x'}$ and $I_{z'y'}$ (first and second subscripts refer to the axes of propagation and polarization, respectively), have contributions of their $I_\parallel$, $I_\perp$, and $I_{ax}$ configurations:[5]

$$I_{z'y'} = I_\parallel cos^2\theta' + I_\perp sin^2\theta' sin^2\varphi' + I_{ax} sin^2\theta' cos^2\varphi' \tag{S1}$$

$$I_{z'x'} = I_\parallel sin^2\theta' sin^2\varphi' + I_\perp cos^2\theta' + I_{ax} sin^2\theta' cos^2\varphi' \tag{S2}$$

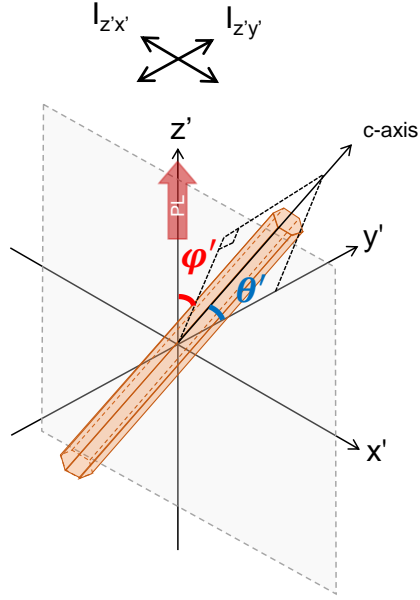

**Figure S10.** The same figure as **Figure 5** in the main text, displayed here again to facilitate the discussions.

**Determination of 3D orientation of a single microcrystal via two magnetic transition peaks.** For magnetic transition peaks, the relation of $I_{ax} = I_\parallel$ transforms **Equation (S1)** and **(S2)** to:

$$I_{z'y'} = I_\perp \sin^2\theta' \sin^2\varphi' + I_\parallel(\cos^2\varphi' + \sin^2\varphi'\cos^2\theta') \tag{S3}$$

$$I_{z'x'} = I_\parallel \sin^2\theta' + I_\perp \cos^2\theta' \tag{S4}$$

in which we need to measure $(I_{z'y'}, I_{z'x'})$ and $(I_\parallel, I_\perp)$ to determine $(\theta', \varphi')$. Although $(I_{z'y'}, I_{z'x'})$ is ready to measure, measuring $(I_\parallel, I_\perp)$ requires to put the same microcrystal again to the in-plane configuration on the substrate to allow detection normal to the c-axis, which is not practical. Base on this point, we develop a more sophisticated method to use the value of $I_\parallel/I_\perp$ instead of $(I_\parallel, I_\perp)$. Since the $I_\parallel/I_\perp$ is directly linked to the LDOP which is identical for the same type of microcrystals, it can be viewed as a known parameter derived from earlier in-plane spectroscopy on microcrystals of the same type other than the one measured here. By considering the number of the unknown parameters in the **Equation (S3)** and **(S4)**, we need, at least, two magnetic transition peaks with their respective $I_\parallel/I_\perp$ to calculate the $(\theta', \varphi')$.

$$\text{MD Peak 1:} \begin{cases} \dfrac{I_{1\parallel}}{I_{1\perp}} = A' \\ I_{1z'x'} = I_{1\perp}\cos^2\theta' + I_{1\parallel}\sin^2\theta' \\ I_{1z'y'} = I_{1\perp}\sin^2\theta'\sin^2\varphi' + I_{1\parallel}(\cos^2\varphi' + \sin^2\varphi'\cos^2\theta') \end{cases} \tag{S5}$$

MD Peak 2:
$$\begin{cases} \frac{I_{2\|}}{I_{2\perp}} = B' \\ I_{2z'x'} = I_{2\perp}\cos^2\theta' + I_{2\|}\sin^2\theta' \\ I_{2z'y'} = I_{2\perp}\sin^2\theta'\sin^2\varphi' + I_{2\|}(\cos^2\varphi' + \sin^2\varphi'\cos^2\theta') \end{cases} \quad (S6)$$

Where A' and B' are constants determined by the linear polarization degrees (LDOPs) of the two selected magnetic transition peaks, and $I_{nz'x'}$ and $I_{nz'y'}$ (n=1 or 2) indicate the collected emission intensities of the two transition peaks at two orthogonal polarization angles, respectively. Combining the **Equation (S5)** and **(S6)** yields the expression of $(\theta', \varphi')$ as:

$$\sin^2\theta' = \frac{\left(\frac{I_{2z'y'}}{I_{2z'x'}}-B'\right)(1-A')-\left(\frac{I_{1z'y'}}{I_{1z'x'}}-A'\right)(1-B')}{\left(\frac{I_{2z'y'}}{I_{2z'x'}}-\frac{I_{1z'y'}}{I_{1z'x'}}\right)(A'-1)(B'-1)} \quad (S7)$$

$$\begin{cases} \sin^2\varphi' = \dfrac{\frac{I_{1z'y'}}{I_{1z'x'}}[\sin^2\theta'(A'-1)+1]-A'}{\sin^2\theta'(1-A')} \\ \sin^2\varphi' = \dfrac{\frac{I_{2z'y'}}{I_{2z'x'}}[\sin^2\theta'(B'-1)+1]-B'}{\sin^2\theta'(1-B')} \end{cases} \quad (S8)$$

It is clear that the 3D information $(\theta', \varphi')$ can be calculated using two magnetic transition peaks at two orthogonal polarization detections.

Discussions:

(1) When A' or B' = 1 (i.e. any of the two selected peaks is not polarized), there is no solution of the **Equation (S7)** and **(S8)**.

(2) When A' = B' (i.e. the two selected peaks have the same polarization degree), there is no solution of the **Equation (S7)** and **(S8)**.

In summary, the 3D orientation of a single microcrystal can be effectively determined by a set of orthogonal polarization detections in using two *polarized* magnetic transition peaks with the *unequal* polarization degrees.

**Determination of 3D orientation of a single microcrystal via two electric transition peaks.** For electric transition peaks, the relation of $I_{ax} = I_\perp$ transforms **Equation (S1)** and **(S2)** to:

$$I_{z'y'} = I_\|\cos^2\theta' + I_\perp\sin^2\theta' \quad (S9)$$

$$I_{z'x'} = I_\|\sin^2\theta'\sin^2\varphi' + I_\perp(\cos^2\varphi' + \sin^2\varphi'\cos^2\theta') \quad (S10)$$

Also, we need at least two electric transition peaks with their respective $I_\|/I_\perp$ to calculate the $(\theta', \varphi')$ like what we have done for two magnetic peaks.

ED Peak 1:
$$\begin{cases} \frac{I_{1\|}}{I_{1\perp}} = C' \\ I_{1z'y'} = I_{1\|}\cos^2\theta' + I_{1\perp}\sin^2\theta' \\ I_{1z'x'} = I_{1\|}\sin^2\theta'\sin^2\varphi' + I_{1\perp}(\cos^2\varphi' + \sin^2\varphi'\cos^2\theta') \end{cases} \quad (S11)$$

ED Peak 2: 
$$\begin{cases} \frac{I_{2\parallel}}{I_{2\perp}} = D' \\ I_{2z'y'} = I_{2\parallel}\cos^2\theta' + I_{2\perp}\sin^2\theta' \\ I_{2z'x'} = I_{2\parallel}\sin^2\theta'\sin^2\varphi' + I_{2\perp}(\cos^2\varphi' + \sin^2\varphi'\cos^2\theta') \end{cases}$$ (S12)

Then, we can obtain the expressions of $(\theta', \varphi')$ by combining the **Equation (S11)** and **(S12)**.

$$\sin^2\theta' = \frac{\left(\frac{I_{2z'x'}}{I_{2z'y'}}D'-1\right)(C'-1)-\left(\frac{I_{1z'x'}}{I_{1z'y'}}C'-1\right)(D'-1)}{\left(\frac{I_{2z'x'}}{I_{2z'y'}}-\frac{I_{1z'x'}}{I_{1z'y'}}\right)(C'-1)(D'-1)}$$ (S13)

$$\begin{cases} \sin^2\varphi' = \frac{\frac{I_{1z'x'}}{I_{1z'y'}}[C'+(1-C')\sin^2\theta']-1}{\sin^2\theta'(C'-1)} \\ \sin^2\varphi' = \frac{\frac{I_{2z'x'}}{I_{2z'y'}}[D'+(1-D')\sin^2\theta']-1}{\sin^2\theta'(D'-1)} \end{cases}$$ (S14)

Discussions:

(3) When C' or D' = 1 (i.e. any of the two selected peaks is not polarized), there is no solution of the **Equation (S13)** and **(S14)**.

(4) When C' = D' (i.e. the two selected peaks have the same polarization degree), there is no solution of the **Equation (S13)** and **(S14)**.

In summary, the 3D orientation of a single microcrystal can be effectively determined by a set of orthogonal polarization detections in using two *polarized* electric transition peaks with the *unequal* polarization degrees.

**Determination of 3D orientation of a single microcrystal via one magnetic and one electric transition peaks.** We can also use one magnetic and one electric transition peaks to determine the 3D information $(\theta', \varphi')$ of a single microcrystal.

One MD Peak: 
$$\begin{cases} \frac{I_{1\parallel}}{I_{1\perp}} = E' \\ I_{1z'x'} = I_{1\perp}\cos^2\theta' + I_{1\parallel}\sin^2\theta' \\ I_{1z'y'} = I_{1\perp}\sin^2\theta'\sin^2\varphi' + I_{1\parallel}(\cos^2\varphi' + \sin^2\varphi'\cos^2\theta') \end{cases}$$ (S15)

One ED Peak: 
$$\begin{cases} \frac{I_{2\parallel}}{I_{2\perp}} = F' \\ I_{2z'y'} = I_{2\parallel}\cos^2\theta' + I_{2\perp}\sin^2\theta' \\ I_{2z'x'} = I_{2\parallel}\sin^2\theta'\sin^2\varphi' + I_{2\perp}(\cos^2\varphi' + \sin^2\varphi'\cos^2\theta') \end{cases}$$ (S16)

From the **Equation (S15)** and **(S16)** the $(\theta', \varphi')$ can be expressed as:

$$\sin^2\theta' = \frac{\frac{I_{1z'y'}}{I_{1z'x'}}\cdot\frac{I_{2z'y'}}{I_{2z'x'}}(F'-1)+\frac{I_{2z'y'}}{I_{2z'x'}}(1-E'F')+E'F'-F'}{\left(1-\frac{I_{1z'y'}}{I_{1z'x'}}\cdot\frac{I_{2z'y'}}{I_{2z'x'}}\right)(E'-1)(F'-1)}$$ (S17)

$$\begin{cases} \sin^2\varphi' = \dfrac{\dfrac{I_{1z'y'}}{I_{1z'x'}}(\cos^2\theta' + E'\sin^2\theta') - E'}{\sin^2\theta'(1-E')} \\ \sin^2\varphi' = \dfrac{F'\cos^2\theta' + \sin^2\theta' - \dfrac{I_{2z'y'}}{I_{2z'x'}}}{\dfrac{I_{2z'y'}}{I_{2z'x'}}\sin^2\theta'(F'-1)} \end{cases} \quad (S18)$$

Where E' and F' are determined by the linear polarization degrees (LDOPs) of the selected magnetic and electric transition peaks, and $I_{nz'x'}$ and $I_{nz'y'}$ indicate the collected emission intensities of the two transition peaks (n=1 refers to MD or n=2 to ED) at two orthogonal polarization angles, respectively.

Discussions:

(5) When E' or F' = 1 (i.e. any of the two selected peaks is not polarized), there is no solution of the **Equation (S17)** and **(S18)**.

In summary, the 3D orientation of a single microcrystal can be effectively determined by a set of orthogonal polarization detections in using one *polarized* electric and one *polarized* magnetic transition peaks *even if* they have the same linear polarization degree (LDOP).